\documentstyle[twoside,fleqn,espcrc2]{article}

\newcommand{\AmS}{{\protect\the\textfont2
  A\kern-.1667em\lower.5ex\hbox{M}\kern-.125emS}}

\title{ A new algorithm of Langevin simulation \\
        and its application to 
    the SU(2) and SU(3) lattice gauge 
  \thanks{Presented by S. Furui }}
\author{Hideo Nakajima\address{Department of Information Science, 
Utsunomiya University,\\
2753 Ishii, Utsunomiya 321 Japan }
\thanks{e-mail nakajima@kinu.infor.utsunomiya-u.ac.jp}
and 
Sadataka Furui \address{School of Science and Engineering, 
Teikyo University, \\
1-1 Toyosatodai, Utsunomiya 320, Japan}
\thanks{e-mail furui@dream.ics.teikyo-u.ac.jp}
}

\begin{document}

\begin{abstract}
The 2nd order Runge-Kutta scheme Langevin simulation of unquenched QCD 
in pseudofermion method derived from our general theory shows a behaviour
as a function of the Langevin step t better than the Fukugita,Oyanagi,Ukawa's
 scheme.
\end{abstract}

\maketitle

\section{Introduction} 

In order to incorporate fermions in the lattice QCD, Hybrid Monte Carlo(HMC)
is the standard method\cite{Gupta}, but in the case of large lattice, the
method is not free from problems caused by rounding errors\cite{jan}. The 
Langevin simulation is another possible method to simulate fields including 
fermions with or without gauge fixing.  

The discretized Langevin simulation results depend on the step size $t$, and
the incorporation of higher order corrections in the step function
to reproduce the Fokker-Planck distribution is the problem.
Drummond et al\cite{DDH} showed a method to derive higher order corrections
up to second order in the step in  U(1) spin model and in SU(N) lattice gauge 
theory. We obtained the Langevin step function up to third order in 
flat space and up to second order in curved space\cite{NF}. 

A Langevin simulation using the Runge-Kutta type algorithm including dynamical
 quark loops and corrections up to second order in $t$ was proposed
by Batrouni et al\cite{BKKLSW} and, Ukawa and Fukugita\cite{UF}.

An extensive full-QCD Langevin simulation was performed by Fukugita, Oyanagi 
and Ukawa\cite{FOU} in partial 2nd order bilinear noise scheme. They reported 
that the 2nd order pseudo-fermion scheme is worse than the partial 2nd order
bilinear noise scheme in a simulation of $4^3\times 8$ lattice.

In this work, we extend the new algorithm of Langevin simulation 
to an algorithm in the canonical coordinate space of Lie algebra of
SU(N). We then apply the method to SU(N) lattice gauge theory including
pseudofermion and discuss possible improvements.

\section{Random walks in the space of Lie groups}

In the space of SU(N) Lie groups, a group element $g$ is expressed as 
$g=\exp(x\cdot \lambda)$ 
where $x_i$'s are the canonical coordinates of the first kind, and
$\lambda_i$'s are antihermitean matrices which span space of $su(N)$ Lie algebra as $[\lambda_i, \lambda_j]=c_{ijk}\lambda_k $
and are normalized as
$Tr(\lambda_i^{\dag} \lambda_j)=\delta_{ij}$ .

The right differentiation $\nabla$ is defined for any $f(g)$ as
$e^{ X\cdot\nabla} f(g)=f(g e^{ X\cdot \lambda})$ ,
where $\nabla_i$'s follow the same commutation relation as
$\lambda_i$'s and 
\begin{equation}
e^{- X\cdot\nabla} \delta^{inv}(g -g_0)=\delta^{inv}(g -g_0e^{ X\cdot \lambda})
\end{equation}

The Fokker-Planck equation in the space of Lie group is  
$\partial_t \phi=K\phi$ ,
where
$K=\nabla_k(\nabla_k-u_k)$ ,
and $\phi$ is normalized with the Haar measure.

If $u_i=\nabla_i S$, then $\phi$ tends to its asymptotic distribution
$const \cdot e^S$.
The formal solution $\phi$ with the initial condition $\phi(g)=\delta^{inv}
(g-g_0)$ is
\begin{equation}
\phi=e^{tK}\delta^{inv}(g-g_0).
\end{equation}

The Langevin simulation algorithm $X_i$ for the random walk in the
Lie algebra space is defined as
\begin{equation}
e^{tK}\delta^{inv}(g-g_0)=\langle \delta^{inv}(g-g_0e^{X\cdot\lambda})\rangle,
\end{equation}
where $X$ is a function determined from $s=\sqrt{t}$, and all values
of $u_i$ together with
their higher derivatives evaluated at $g_0$, and independent Gaussian random
variates $\xi$'s with variance 2 and mean 0, and $\langle ...\rangle$ denotes an averaging over the Gaussian
distribution.

Thus the problem is to find the $X_i$ such that
\begin{equation}
\langle g|\exp( t\hat K)|g_0\rangle=\langle g|\langle e^{-X(s,\xi)\cdot \nabla}\rangle|g_0\rangle
\end{equation}
where $\hat K=\hat\nabla \cdot (\hat \nabla- u(\hat g))$, 
and note that operators are normal ordered, i.e., all $\hat \nabla$'s are
set in the leftmost position relative to all $\hat g$'s on the l.h.s. of
the equation.

In order to get $X$, let $X_i$ be expanded as
\begin{equation}
X_i=\sum_{n=0}^{\infty} \sum_{k=0}^{n} s^n f_{(n,k)}h_k(\xi)\ \ .
\end{equation}
where $h_k(\xi)$ is a generalized $k$-th order Hermite polynomial.
First, put $\exp(t\hat K)$ in normal order as mentioned above.
Then starting from $X_i=s\xi_i + \cdots$, and considering operator
equations for each order $O(s^n)$, and then for given $n$, $O(\nabla^m)$
in larger-$m$-first order, we proceed perturbatively and obtain $f_{(n,k)}$'s.
We can prove this procedure works satisfactorily, i.e., a newly fixed
$f_{(n,k)}$ does not affect lower order equations.

Our $O(s^6)$ algorithm is as follows,
\begin{eqnarray}
&&X_i=s\xi_i+s^2u_i+{s^3\over 12}(6\xi_j \nabla_j u_i-N\xi_i)\nonumber\\
&&+{s^4\over 6}(Nu_i+3u_j \nabla_j u_i+\nabla^2 u_i
+\xi_j \xi_k \nabla_j \nabla_k u_i)\nonumber\\
&&+{s^5\over 288}(12c_{ijk}c_{klm}\xi_l u_j u_m
-36N\xi_j \nabla_j u_i \nonumber\\
&&      + 12\xi_j \nabla_k u_j \nabla_k u_i
+ 48\xi_j \nabla_j u_k \nabla_k u_i\nonumber\\
&&      + 48\xi_j u_k \nabla_k \nabla_j u_i
+ 48\xi_j u_k \nabla_j \nabla_k u_i\nonumber\\
&&      + 96\xi_j \nabla_j \nabla^2 u_i
+ 24c_{ijk}c_{ljm}\xi_l \nabla_k u_m\nonumber\\
&&      + 24c_{ijk}\xi_l u_j \nabla_l u_k
+ 48c_{ijk}\xi_l \nabla_j \nabla_l u_k\nonumber\\
&&      - 24c_{jkl}\xi_j u_k \nabla_l u_i
+ 48c_{jkl}\xi_j \nabla_k \nabla_l u_i
- N^2 \xi_i)\nonumber\\
&&+{s^6 \over 60}(-30N u_j \nabla_j u_i
-10\nabla^2 u_j \nabla_j u_i\nonumber\\
&&      -10 u_j \nabla_j u_k \nabla_k u_i
-30 N \nabla^2 u_i\nonumber\\
&&      -20 \nabla_j u_k \nabla_j \nabla_k u_i
-10 u_j u_k \nabla_k \nabla_j u_i\nonumber\\
&&      -20 u_j \nabla_j \nabla^2 u_i
-10 \nabla^2 \nabla^2 u_i
 -10 c_{ijk} \nabla_j u_l \nabla_l u_k\nonumber\\
&&      -5 c_{ijk} u_j \nabla^2 u_k
-5 c_{ijk} u_j u_l \nabla_l u_k\nonumber\\
&&      -5 c_{ijk}c_{ljm}u_l \nabla_m u_k-5 c_{ijk}c_{lkm}u_l \nabla_j u_m
\nonumber\\
&&      -54 N^2 u_i)
\end{eqnarray}

In the case of SU(3), exponentiation of the matrix $x\cdot\lambda$ is done
analytically as follows. 
\begin{equation}
e^{x\cdot \lambda}=e^{i z_1}P_1+e^{i z_2}P_2+e^{i z_3}P_3.
\end{equation}
with $z_i$ as the eigenvalue given by the Cardano method, and $P_i$ as 
the projection operator on its eigenspace.
The unitarity of the matrix is preserved in very high accuracy.

\subsection{QCD with pseudofermion}

In order to incorporate quark loops in the vacuum, we include in addition to
 the gauge field $U$, a pseudofermion field $\phi$ which is a complex-scalar, 
and define the partition $Z=\int d\phi^* d\phi dU e^{\beta S}$.

The action is
\begin{equation}
S=S_{gauge}+S_{fermi}
\end{equation}
where
\begin{equation}
S_{gauge}={1 \over N}\sum_{p}Re(TrU_{p})
\end{equation}

The matrix elements for Wilson fermions are calculated by using hermitian
$\gamma$ matrices and the hopping parameter $\kappa$.

After the preconditionings the fermionic action can be replaced by the
even-site fermionic action\cite{dGR}
\begin{equation}
S_{fermi}=-\phi_e^* {1\over {\tilde M \tilde M^\dagger}}\phi_e
\end{equation}
where $\tilde M=1-\kappa^2 D D$ and  
$D_{xy}=\sum_\mu\{ \\
(1-\gamma_\mu)U_{x,\mu}\delta_{x+\mu,y}
     +(1+\gamma_\mu)U_{y,\mu}^\dagger\delta_{x-\mu,y}\}$.

The $O(s^4)$ Runge-Kutta algorithm $X$ for the SU(N) link variables U is
\begin{equation}
 X={N\over 12}s^2(-s \xi+s^2 2 u)+s \xi+{1\over 2}s^2(u +\tilde u)
\end{equation}
where $\tilde u=u(g e^{(s \xi+s^2 u)\cdot \lambda} )$.
The velocity field is
\begin{equation}
u_{i\mu}=\nabla^i_{\mu} S_{gauge}+\nabla^i_{\mu} S_{fermi}.
\end{equation}
where, $\nabla^i_{\mu}$ is the right differentiation with respect to the i-th
canonical coordinate of $U_{x,\mu}$. 

The function $X$ for the pseudofermion is similar but the
$N$ dependent term is absent,  $u_\phi=-2{1\over \tilde M\tilde M^\dagger}\phi$ and $u_{\phi^*}=-2\phi^*{1\over \tilde M\tilde M^\dagger}$.

\section{Numerical Results}

In the second order Taylor scheme Langevin simulation of SU(2) on $8^4$ lattice, we found the dependence on $t$ of the data 
simulated in the space of Lie algebra is much better than that in the space of ${\bf S}^3$\cite{kyoto}. 

Simulations of SU(2) single degree of freedom, and of $4^4$ and $8^4$ lattices
were performed in the Taylor scheme and in the Runge-Kutta scheme. 
In the single degree of freedom both showed similar dependence, however in the
 $4^4$ lattice, the behaviour of the Taylor scheme 
was worse than the Runge-Kutta scheme at large $ t$\cite{kyoto}.  

In the case of SU(2) single degree of freedom, the third order Taylor scheme 
did not improve the second order scheme, but Runge-Kutta scheme that
incorporates the third order term is expected to improve the dependence on
$t$ in the lattice simulation. 

We checked the results of SU(3) by comparing with the Monte Carlo simulation, 
in the Cabibbo-Marinari\cite{CM}'s scheme. 
The dependence of $\langle U_P\rangle=\langle Re({1\over 3}Tr U_p)\rangle$  
on $\kappa$ in the $4^4$ lattice in the 2nd order Langevin simulation of 
full-QCD with pseudofermion is shown in Fig.1. At $t=0.05$, the agreement to 
the HMC result is better than that of Fukugita, Oyanagi and Ukawa\cite{FOU}
and at $ t=0.01$ they are consistent within statistical errors.

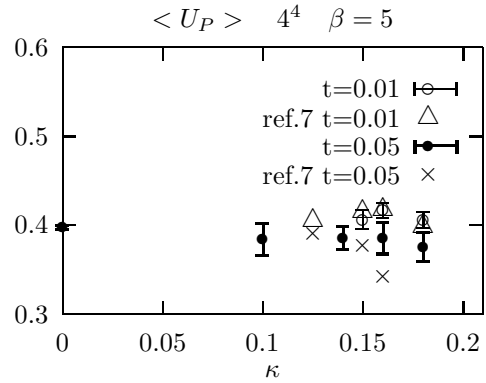
\begin{figure}
\setlength{\unitlength}{0.240900pt}
\ifx\plotpoint\undefined\newsavebox{\plotpoint}\fi
\sbox{\plotpoint}{\rule[-0.200pt]{0.400pt}{0.400pt}}%
\begin{picture}(900,600)(0,0)
\sbox{\plotpoint}{\rule[-0.200pt]{0.400pt}{0.400pt}}%
\put(176.0,113.0){\rule[-0.200pt]{0.400pt}{100.937pt}}
\put(176.0,113.0){\rule[-0.200pt]{4.818pt}{0.400pt}}
\put(154,113){\makebox(0,0)[r]{0.3}}
\put(816.0,113.0){\rule[-0.200pt]{4.818pt}{0.400pt}}
\put(176.0,253.0){\rule[-0.200pt]{4.818pt}{0.400pt}}
\put(154,253){\makebox(0,0)[r]{0.4}}
\put(816.0,253.0){\rule[-0.200pt]{4.818pt}{0.400pt}}
\put(176.0,392.0){\rule[-0.200pt]{4.818pt}{0.400pt}}
\put(154,392){\makebox(0,0)[r]{0.5}}
\put(816.0,392.0){\rule[-0.200pt]{4.818pt}{0.400pt}}
\put(176.0,532.0){\rule[-0.200pt]{4.818pt}{0.400pt}}
\put(154,532){\makebox(0,0)[r]{0.6}}
\put(816.0,532.0){\rule[-0.200pt]{4.818pt}{0.400pt}}
\put(176.0,113.0){\rule[-0.200pt]{0.400pt}{4.818pt}}
\put(176,68){\makebox(0,0){0}}
\put(176.0,512.0){\rule[-0.200pt]{0.400pt}{4.818pt}}
\put(333.0,113.0){\rule[-0.200pt]{0.400pt}{4.818pt}}
\put(333,68){\makebox(0,0){0.05}}
\put(333.0,512.0){\rule[-0.200pt]{0.400pt}{4.818pt}}
\put(490.0,113.0){\rule[-0.200pt]{0.400pt}{4.818pt}}
\put(490,68){\makebox(0,0){0.1}}
\put(490.0,512.0){\rule[-0.200pt]{0.400pt}{4.818pt}}
\put(647.0,113.0){\rule[-0.200pt]{0.400pt}{4.818pt}}
\put(647,68){\makebox(0,0){0.15}}
\put(647.0,512.0){\rule[-0.200pt]{0.400pt}{4.818pt}}
\put(805.0,113.0){\rule[-0.200pt]{0.400pt}{4.818pt}}
\put(805,68){\makebox(0,0){0.2}}
\put(805.0,512.0){\rule[-0.200pt]{0.400pt}{4.818pt}}
\put(176.0,113.0){\rule[-0.200pt]{158.994pt}{0.400pt}}
\put(836.0,113.0){\rule[-0.200pt]{0.400pt}{100.937pt}}
\put(176.0,532.0){\rule[-0.200pt]{158.994pt}{0.400pt}}
\put(506,23){\makebox(0,0){$\kappa$}}
\put(506,577){\makebox(0,0){$< U_P >\quad  4^4\quad  \beta=5$}}
\put(176.0,113.0){\rule[-0.200pt]{0.400pt}{100.937pt}}
\put(706,467){\makebox(0,0)[r]{t=0.01}}
\put(750,467){\circle{18}}
\put(647,261){\circle{18}}
\put(679,276){\circle{18}}
\put(742,260){\circle{18}}
\put(728.0,467.0){\rule[-0.200pt]{15.899pt}{0.400pt}}
\put(728.0,457.0){\rule[-0.200pt]{0.400pt}{4.818pt}}
\put(794.0,457.0){\rule[-0.200pt]{0.400pt}{4.818pt}}
\put(647.0,246.0){\rule[-0.200pt]{0.400pt}{7.227pt}}
\put(637.0,246.0){\rule[-0.200pt]{4.818pt}{0.400pt}}
\put(637.0,276.0){\rule[-0.200pt]{4.818pt}{0.400pt}}
\put(679.0,264.0){\rule[-0.200pt]{0.400pt}{5.541pt}}
\put(669.0,264.0){\rule[-0.200pt]{4.818pt}{0.400pt}}
\put(669.0,287.0){\rule[-0.200pt]{4.818pt}{0.400pt}}
\put(742.0,248.0){\rule[-0.200pt]{0.400pt}{6.022pt}}
\put(732.0,248.0){\rule[-0.200pt]{4.818pt}{0.400pt}}
\put(732.0,273.0){\rule[-0.200pt]{4.818pt}{0.400pt}}
\put(706,422){\makebox(0,0)[r]{ref.7 t=0.01}}
\put(750,422){\makebox(0,0){$\triangle$}}
\put(569,260){\makebox(0,0){$\triangle$}}
\put(647,274){\makebox(0,0){$\triangle$}}
\put(679,278){\makebox(0,0){$\triangle$}}
\put(742,250){\makebox(0,0){$\triangle$}}
\sbox{\plotpoint}{\rule[-0.400pt]{0.800pt}{0.800pt}}%
\put(706,377){\makebox(0,0)[r]{t=0.05}}
\put(750,377){\circle*{18}}
\put(176,249){\circle*{18}}
\put(490,230){\circle*{18}}
\put(616,232){\circle*{18}}
\put(679,232){\circle*{18}}
\put(742,218){\circle*{18}}
\put(728.0,377.0){\rule[-0.400pt]{15.899pt}{0.800pt}}
\put(728.0,367.0){\rule[-0.400pt]{0.800pt}{4.818pt}}
\put(794.0,367.0){\rule[-0.400pt]{0.800pt}{4.818pt}}
\put(176.0,246.0){\rule[-0.400pt]{0.800pt}{1.445pt}}
\put(166.0,246.0){\rule[-0.400pt]{4.818pt}{0.800pt}}
\put(166.0,252.0){\rule[-0.400pt]{4.818pt}{0.800pt}}
\put(490.0,205.0){\rule[-0.400pt]{0.800pt}{12.045pt}}
\put(480.0,205.0){\rule[-0.400pt]{4.818pt}{0.800pt}}
\put(480.0,255.0){\rule[-0.400pt]{4.818pt}{0.800pt}}
\put(616.0,214.0){\rule[-0.400pt]{0.800pt}{8.672pt}}
\put(606.0,214.0){\rule[-0.400pt]{4.818pt}{0.800pt}}
\put(606.0,250.0){\rule[-0.400pt]{4.818pt}{0.800pt}}
\put(679.0,207.0){\rule[-0.400pt]{0.800pt}{12.045pt}}
\put(669.0,207.0){\rule[-0.400pt]{4.818pt}{0.800pt}}
\put(669.0,257.0){\rule[-0.400pt]{4.818pt}{0.800pt}}
\put(742.0,195.0){\rule[-0.400pt]{0.800pt}{11.081pt}}
\put(732.0,195.0){\rule[-0.400pt]{4.818pt}{0.800pt}}
\put(732.0,241.0){\rule[-0.400pt]{4.818pt}{0.800pt}}
\sbox{\plotpoint}{\rule[-0.200pt]{0.400pt}{0.400pt}}%
\put(706,332){\makebox(0,0)[r]{ref.7 t=0.05}}
\put(750,332){\makebox(0,0){$\times$}}
\put(569,240){\makebox(0,0){$\times$}}
\put(647,221){\makebox(0,0){$\times$}}
\put(679,172){\makebox(0,0){$\times$}}
\end{picture}
\caption{Dependence of Wilson loop $U_P=Re({1\over 3}Tr U_p)$ on the hopping parameter $\kappa$. 
The triangle at $\kappa=0.15$ is the result of HMC[1]. }
\end{figure}
\section{Outlook} 

General algorithm for obtaining higher order weak Taylor approximations to the Langevin
step algorithms with respect to $t$, improved version in practical use
for a finite fixed order was formulated. (cf. Nakajima \cite{N}
for full order weak Taylor algorithm in recursion formula) 

We try to obtain a local $O(t^3)$ Runge-Kutta like algorithm in $SU(3)$ 
and aim to use it for investigation of its performance in full QCD simulation.

This work was supported by National Laboratory for High Energy Physics, as
KEK Supercomputer Project ( Project No.97-23 ).

\end{document}